\documentclass[aps,pre,twocolumn,groupedaddress,superscriptaddress]{revtex4-1}
\usepackage{amsmath,amsbsy,amssymb, graphicx, textcomp, upgreek, hyperref, subfigure}

\usepackage[dvipsnames]{xcolor} 

\usepackage{todonotes}

\usepackage{tikz}
\usetikzlibrary{calc}
\usetikzlibrary{fpu}
\usetikzlibrary{arrows}
\usetikzlibrary{decorations,decorations.shapes}

\makeatletter
\newcommand*{\balancecolsandclearpage}{%
  \close@column@grid
  \clearpage
  \twocolumngrid
}
\makeatother

\begin{document}

\title{Energy harvesting through gas dynamics in the free molecular flow regime between structured surfaces at different temperatures.}

\author{Tobias Baier}
\email[]{baier@csi.tu-darmstadt.de}
\affiliation{Center of Smart Interfaces, TU Darmstadt, Germany}

\author{Julia D\"olger}
\affiliation{Department of Physics, Technical University of Denmark, Denmark}

\author{Steffen Hardt}
\email[]{hardt@csi.tu-darmstadt.de}
\affiliation{Center of Smart Interfaces, TU Darmstadt, Germany}

\date{\today}

\begin{abstract}
For a gas confined between surfaces held at different temperatures the velocity distribution shows a significant deviation from the Maxwell distribution when the mean free path of the molecules is comparable to or larger than the channel dimensions. If one of the surfaces is suitably structured, this non-equilibrium distribution can be exploited for momentum transfer in tangential direction between the two surfaces. This opens up the possibility to extract work from the system which operates as a heat engine. Since both surfaces are held at constant temperatures, the mode of momentum transfer is different from thermal creep flow that has gained more attention so far. This situation is studied in the limit of free-molecular flow for the case that an unstructured surface is allowed to move tangentially with respect to a structured surface. Parameter studies are conducted, and configurations with maximum thermodynamic efficiency are identified. Overall, it is shown that significant efficiencies can be obtained by tangential momentum transfer between structured surfaces.  
\end{abstract}

\pacs{47.61.-k, 47.45.Dt, 05.60.-k, 44.15.+a}


\maketitle

\section{Introduction\label{sec:intro}}

As the length scales of many technological devices have shrunk to the order of the mean free path of gas molecules at standard conditions, transport phenomena occurring in the transition flow or free-molecular flow regime have gained increased interest, particularly within the research community concerned with micro- and nanosystems. For such systems it is no longer possible to describe transport phenomena by the usual continuum models such as the Navier-Stokes equations, but the Boltzmann equation has to be employed to capture the physics \cite{Sone_2007}. Along with such a scenario come a number of effects that are absent in gases within the continuum regime. As an example, flows that are induced by a temperature gradient appear. Such thermally induced gas flows have been exploited already a long time ago, for example in the Crookes radiometer \cite{Crookes_1876} or in Knudsen pumps \cite{Knudsen_1910}. In these setups, the Knudsen number was increased by rarefaction occurring at reduced pressures. At the time these studies were conducted the molecular picture of matter was still debated. Nevertheless, the basic theoretical framework for rarefied gas dynamics had already been put forward \cite{Maxwell_1860, Boltzmann_1896}, and corresponding experiments triggered the further development of models connecting continuum mechanics with gas kinetics \cite{Chapman_1970}.

Variants of these classical configurations remain active topics of research today \cite{Gupta_2012, Donkov_2011, Taguchi_2010, McNamara_2005, Muntz_2002, Sone_1996}. Depending on the exact form of the thermal and geometric boundary conditions, such thermally induced flows are termed, for example, thermal creep, thermal stress slip or thermal edge flows \cite{Sone_2007}. In such situations, when a net flow is induced within the gas, momentum conservation dictates that a net force is exerted onto the solid forming a boundary to the flow, as evidenced by the rotation of the Crookes radiometer \footnote{Note, however, that in the limit of infinite $\mathrm{Kn}$ net mass and net momentum flux is not necessarily coupled \citep{Sone_2007, Donkov_2011}.}. That way it is possible to convert thermal into mechanical energy, i.e. to build a heat engine based on this principle.

\begin{figure}
\begin{tikzpicture}[scale=3/4, >=stealth]
  \def\angAlpha{25}
  \def\angBeta{15}
  \def\lenOne{8} 
  \def\lenGap{1.5} 

  \pgfmathparse{\lenOne*sin(\angAlpha)/sin(90-\angAlpha+\angBeta)}  
  \def\lenTwo{\pgfmathresult} 

  \coordinate (t1) at ($(0:1)$);   
  \coordinate (t2) at ($(-90+\angBeta:1)$);
  \coordinate (t3) at ($(\angAlpha:1)$);
  
  \coordinate (n1) at ($(-90:1)$);
  \coordinate (n2) at ($(\angBeta:1)$);
  \coordinate (n3) at ($(90+\angAlpha:1)$);

  \coordinate (a) at (0,0);
  \coordinate (b) at ($(a) +\lenTwo*(t2)$);
  \coordinate (d) at ($(a) +\lenOne*(t1)$);
  \coordinate (e) at ($(a) -\lenGap*(n1)$);
  \coordinate (f) at ($(d) -\lenGap*(n1)$);
  
  \coordinate (m1) at ($(a)+2/5*(d)-2/5*(a)$);   
  \coordinate (m2) at ($(a)+1/2*(b)-1/2*(a)$);

  \draw[very thick,color=BrickRed] (a) -- node[black,below left] {$2$} (b);		
  \draw[thick,blue,dashed] (a) -- (d);																
  \draw[very thick,gray] (b) -- node[black,below]{$3$} (d);								
  \draw[very thick,blue] (e) -- node[black,above] {$1$} (f);								
  \draw[very thick,gray,dashed] (a) -- node[black,left] {$P$} (e);
  \draw[very thick,gray,dashed] (d) -- node[black,right] {$P'$}  (f);
  
  \draw[<->, very thick] ($(a)-1.*(t1)$) -- node[black,left] {$H$} ($(e)-1.*(t1)$);
  \draw[<->, very thick] ($(e)-1.*(n1)$) -- node[black,above] {$L$} ($(f)-1.*(n1)$);
  
  \draw[dashed] (a) -- +(-90:1.6);
  \pgfmathparse{-90+\angBeta)}  																		
  \draw ($(a)+(-90:1.4)$) node[left] {$\gamma$} arc (-90:\pgfmathresult:1.4); 		
  \pgfmathparse{-180+\angAlpha)}  																	
  \draw ($(d)+(-180:1.5)$) arc (-180:\pgfmathresult:1.5);													
  \draw ($(d)+(-180+\angAlpha/2:1.8)$) node {$\alpha$};
  
\end{tikzpicture}
\caption{\label{fig:schematic0} (Color online) One segment of the considered periodic domain. Wall 1 and 2 are diffusely reflecting, being  held at temperatures $T_1$ and $T_2$ respectively. Wall 3 is a specularly reflecting surface.  $P$, $P'$ are the periodic boundaries. For extraction of mechanical energy we will assume wall 1 to be able to move in tangential direction.}
\end{figure}
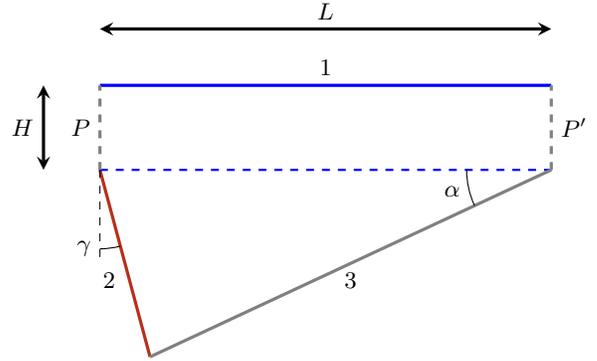

A Knudsen pump is based on a thermal gradient along a narrow channel or slit, for example connected cavities within a porous material or a capillary. Here a gas flow, termed thermal transpiration, is induced in the direction of the temperature gradient. An alternative configuration was considered in \cite{Donkov_2011}, where a 2D channel with structured walls and different temperatures on the two opposing boundaries was studied, c.f. figure \ref{fig:schematic0}. In contrast to conventional Knudsen pumps, such a configuration allows pumping gases in a direction normal to the main direction of the thermal gradient. Moreover, it can be regarded as a heat engine, enabling conversion of thermal into mechanical energy.

The latter aspect is in the focus of the present article, in which we study the thermodynamic efficiency of energy conversion between appropriately structured walls and identify configurations with maximized efficiencies. Finding efficient materials or devices for waste energy recovery is a very active discipline. One of the main research threads in that context aims at improving the performance of thermoelectric materials \cite{Heremans_2008, Boukai_2008, Hochbaum_2008, Poudel_2008, Hsu_2004, Venkatasubramanian_2001} 
which still suffer from low efficiencies. The alternative conversion principle studied in the present article could open a new direction in the field of waste energy recovery. It differs fundamentally from conventional Knudsen pumps in another important aspect: In the limit of infinite Knudsen number no flow, but a momentum transfer occurs \cite{Donkov_2011} which is the cornerstone of energy conversion. In other words, thermal energy can be converted into mechanical energy without any net motion of the gas. 

At this point we refer to the monograph by Sone \cite{Sone_2007}, Sec.~2.5, for a comprehensive summary of exact results obtained for free molecular flow with Maxwell type boundary conditions. It is also worth noting that with alternative boundary conditions of the Cercignani-Lampis type, a net flow is predicted even in the limit of free molecular flow \cite{Kosuge_2011}, contrary to the situation with Maxwell type boundaries.

For energy conversion we assume that in the system sketched in figure \ref{fig:schematic0} the upper wall, labeled 1, is allowed to slide in tangential direction with respect to the structured surface below under the influence of forces exerted by the molecular exchange between them. A realization of such a periodic geometry could be an inner unstructured cylinder rotating within a structured one or an unstructured disc rotating above a structured one, where in both cases we assume the radii of the cylinders or discs to be much larger than the  length of a unit cell $L$.

\begin{figure}
\begin{tikzpicture}[scale=3/4]
  \def\angAlpha{30}
  \def\angBeta{10}
  \def\lenOne{8} 

  \pgfmathparse{\lenOne*sin(\angAlpha)/sin(90-\angAlpha+\angBeta)}  
  \def\lenTwo{\pgfmathresult} 

  \coordinate (t1) at ($(0:1)$);   
  \coordinate (t2) at ($(-90+\angBeta:1)$);
  \coordinate (t3) at ($(-90+2*\angAlpha-\angBeta:1)$);
  \coordinate (t4) at ($(2*\angAlpha:1)$);
  
  \coordinate (n1) at ($(-90:1)$);
  \coordinate (n2) at ($(\angBeta:1)$);
  \coordinate (n3) at ($(2*\angAlpha-\angBeta:1)$);
  \coordinate (n4) at ($(2*\angAlpha+90:1)$);

  \coordinate (a) at (0,0);
  \coordinate (b) at ($(a) +\lenTwo*(t2)$);
  \coordinate (c) at ($(b) +\lenTwo*(t3)$);
  \coordinate (d) at ($(a) +\lenOne*(t1)$);
  
  \coordinate (m1) at ($(a)+2/5*(d)-2/5*(a)$);   
  \coordinate (m2) at ($(a)+1/2*(b)-1/2*(a)$);
  \coordinate (m3) at ($(b)+1/2*(c)-1/2*(b)$);
  \coordinate (m4) at ($(c)+2/5*(d)-2/5*(c)$);

  \draw[very thick,color=BrickRed] (a) -- node[black,left] {$2$} (b);								
  \draw[very thick,color=BrickRed,dashed] (b) -- node[black,below left] {$\bar{2}$} (c);		
  \draw[very thick,blue,dashed] (c) -- node[black,below right] {$\bar{1}$} (d);							
  \draw[very thick,blue] (a) -- node[black,above] {$1$} (d);						
  \draw[very thick,gray] (b) -- node[black,below]{$3$} (d);					
  
  \draw[dashed] (a) -- +(-90:1.6);
  \pgfmathparse{-90+\angBeta)}  																		
  \draw ($(a)+(-90:1.4)$) node[left] {$\gamma$} arc (-90:\pgfmathresult:1.4); 		
  \pgfmathparse{-180+\angAlpha)}  																	
  \draw ($(d)+(-180:1.5)$) arc (-180:\pgfmathresult:1.5);													
  \draw ($(d)+(-180+\angAlpha/2:1.8)$) node {$\alpha$};
  
  \draw[very thick,rounded corners,-stealth'] (a) -- node[black,above] {$t_1$} ($(a)+(t1)$);
  \draw[very thick,-stealth'] (a) -- node[black,right] {$t_2$} ($(a)+(t2)$);
  \draw[very thick,-stealth'] (b) -- node[black,below left] {$t_{\bar{2}}$} ($(b)+(t3)$);
  \draw[very thick,-stealth'] (c) -- node[black,below right] {$t_{\bar{1}}$} ($(c)+(t4)$);

  \draw[very thick,-stealth'] ($(m1)$) -- node[black,left] {$n_1$} ($(m1)+(n1)$);
  \draw[very thick,-stealth'] ($(m2)$) -- node[black,above] {$n_2$} ($(m2)+(n2)$);
  \draw[very thick,-stealth'] ($(m3)$) -- node[black,above left] {$n_{\bar{2}}$} ($(m3)+(n3)$);
  \draw[very thick,-stealth'] ($(m4)$) -- node[black,below left] {$n_{\bar{1}}$} ($(m4)+(n4)$);
\end{tikzpicture}
\caption{\label{fig:schematic} (Color online) The reduced geometry is parametrized by the two angles $\alpha$ and $\gamma$ obeying the constraint $\pi/2 >\alpha>0$ and $\alpha>\gamma>\alpha-\pi/2$, such that the geometry bounded by two walls and their mirror images (with respect to wall 3) constitutes a convex domain. We denote the mirror images of walls 1 and 2 as $\bar{1}$, $\bar{2}$.}
\end{figure}
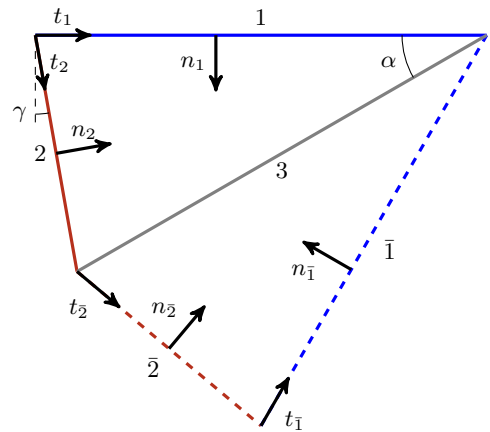

\section{Momentum and energy transfer}\label{sec:transfer}

In what follows, we will assume the Knudsen number to be large enough that the phase space distribution $f(\mathbf{r},\mathbf{c})$ over position $\mathbf{r}$ and velocity $\mathbf{c}$ is governed by the collisionless Boltzmann equation \cite{Sone_2007}, $\mathbf{c} \nabla_\mathbf{r} f(\mathbf{r},\mathbf{c})=0$. If we restrict the analysis to ideally diffuse and specular walls, $f(\mathbf{r},\mathbf{c})$ can thus be found by tracing backwards along $-\mathbf{c}$ till a diffuse boundary is encountered, where the  phase space distribution function is known.

We follow the notation of \cite{Donkov_2011}. In particular, the phase space distribution function for molecules reflected diffusely from a wall at position $\mathbf{r}$ is
\begin{align}
f_r(\mathbf{r},\mathbf{c})&=\nu(\mathbf{r})\mathsf{F}^{2D}(\mathbf{r}, \mathbf{c}), \\
\mathsf{F}^{2D}(\mathbf{r}, \mathbf{c})&=\frac{2}{\sqrt{\pi}}\beta^{3/2} e^{-\beta(\mathbf{c}-\mathbf{u}_{\mathbf{r}})^2}, \quad \beta=\frac{m}{2T_{\mathbf{r}}}, \label{eq:diffusePhaseSpaceDistribution}
\end{align}
where $\nu(\mathbf{r})$ is the particle flux density, i.e. the number of molecules colliding with the wall per unit length and time, $m$ the molecular mass, $T_{\mathbf{r}}$ the wall temperature (in energy units) and $\mathbf{u}_\mathbf{r}$ the velocity of the wall. In this and the following, the term "molecule" is used for the constituents of the gas, even if it may be composed of atoms. The subscript $ \mathbf{r} $ indicates that temperature and velocity are different for different wall segments. The normalization is such that, due to particle number conservation, $\nu(\mathbf{r})=\int_{\mathbf{c}\cdot\mathbf{n}>0}(\mathbf{c}\cdot\mathbf{n})f_r(\mathbf{r},\mathbf{c})d^2\mathbf{c}$, where $\mathbf{n}$ is the inward unit normal vector at the wall. Conversely, the incoming molecular flux is $\nu(\mathbf{r})=-\int_{\mathbf{c}\cdot\mathbf{n}<0}(\mathbf{c}\cdot\mathbf{n})f_i(\mathbf{r},\mathbf{c})d^2\mathbf{c}$, where the inward phase space density $f_i(\mathbf{r},\mathbf{c})=f_r(\mathbf{r}',\mathbf{c}')$ is obtained by tracing backwards along the particle path towards the diffusely reflecting wall at position $\mathbf{r}'$, taking into account each velocity reflection, $\mathbf{c}\to\mathbf{c}'=\mathbf{c}-2\mathbf{c}\mathbf{n}'$, at specularly reflecting walls encountered on the way. The total phase space distribution at a wall is a combination of the inward and reflected distributions, $f(\mathbf{r},\mathbf{c})=\{ f_i(\mathbf{r},\mathbf{c}) \text{ for } \mathbf{c}\cdot\mathbf{n}<0; f_r(\mathbf{r},\mathbf{c}) \text{ for } \mathbf{c}\cdot\mathbf{n}>0 \}$.

In the limiting case of a vanishing gap between the upper and lower surfaces, $H/L=0$, the complexity of the problem is significantly reduced. Since wall 3 is ideally specularly reflective, the backward-tracing procedure can be simplified by considering the original geometry together with its mirror image, as shown in figure \ref{fig:schematic}, where $\bar{1}$ and $\bar{2}$ denote the mirror images of wall 1 and 2, respectively. 
Each wall $i$ is characterized through its tangent and normal vector, $\mathbf{n}_i$ and $\mathbf{t}_i$, its length $l^{(i)}$ and its origin $\mathbf{r}_0^{(i)}$ and can be parametrized by $\mathbf{r}_s^{(i)}=\mathbf{r}_0^{(i)} + s\,l^{(i)}\,\mathbf{t}_i$, $0\le s \le 1$. In the following we describe the positions along the walls by the parameters $s$ and $s'$. The vector joining two positions (from $s'$ to $s$) is $\mathbf{r}_{ss'}=\mathbf{r}_{s'}-\mathbf{r}_{s}$, having length $r_{ss'}$ and normal $\mathbf{n}_{ss'}=\mathbf{r}_{ss'}/r_{ss'}$. Restricting ourselves to a convex domain bounded by walls $1,2,\bar{2},\bar{1}$, the inward particle, momentum and energy fluxes can be expressed as integrals over all other wall segments.

{\it Inward particle, momentum and energy flux:}
\begin{align}\label{eq:influxXi}
\begin{Bmatrix}
\nu(\mathbf{r}_{s}) \\ 
\mathbf{F}_i(\mathbf{r}_{s}) \\ 
\varepsilon_i(\mathbf{r}_{s})
\end{Bmatrix} 
=-\int_{\mathbf{c}\cdot\mathbf{n}<0}\!\!\! (\mathbf{c}\cdot\mathbf{n})f_i(\mathbf{r}_s,\mathbf{c})
\begin{Bmatrix}
1 \\ 
(m\mathbf{c}) \\ 
(\tfrac{1}{2} m\mathbf{c}^2)
\end{Bmatrix} d^2\mathbf{c} &
\\\label{eq:influx}
=\int \! \frac{\cos \vartheta \cos \vartheta'}{r_{ss'}}\nu(\mathbf{r}_{s'}) 
\begin{Bmatrix}
G_{2}(\mathbf{r}_{s'}, \vartheta') \\ 
m \mathbf{n}_{ss'} G_{3}(\mathbf{r}_{s'}, \vartheta') \\ 
\tfrac{1}{2} m G_{4}(\mathbf{r}_{s'}, \vartheta').
\end{Bmatrix} dl_{s'},&
\end{align}
where the integration measure $dl_{s'}$ of the line integral is a shorthand for $|\partial_{s'} \mathbf{r}_{s'}|{ds'}$ with the integration running along all points $\mathbf{r}_{s'}$ on the boundary. Further, $\cos \vartheta = \mathbf{n}_s \mathbf{n}_{s's}$ and $\cos \vartheta' = \mathbf{n}_{s'} \mathbf{n}_{ss'}$ are the cosines of the angles between the connection and the wall normals, and 
\begin{align} \label{eq:gn}
G_{n}(\mathbf{r}_{s'}, \vartheta') &= \int_0^\infty c^n\, \mathsf{F}^{2D}(\mathbf{r}_{s'}, c\,\mathbf{n}_{ss'}) dc
\end{align}
specifies the moments of velocity for molecules emanating from the wall at position $\mathbf{r}_{s'}$ under the angle $\vartheta'$ with respect to the wall normal. This function can be evaluated analytically and is given in appendix \ref{seq:G_n}. Note that it depends implicitly on the wall velocity $\mathbf{u}(\mathbf{r}_{s'})=u(\mathbf{r}_{s'})\mathbf{t}_{s'}$ at position $\mathbf{r}_{s'}$, which we assume to be along the wall tangent $\mathbf{t}_{s'}$ such that the geometry does not change. The angle $\vartheta'$ is specified by $\sin \vartheta' = \mathbf{t}_{s'} \mathbf{n}_{ss'}$, since by construction $\cos\vartheta'\geq 0$ due to the convexity of the domain. We stress that a genuine 2D situation with phase space distribution (\ref{eq:diffusePhaseSpaceDistribution}) is considered; compared to a quasi-2D situation, where the third velocity component has been integrated out, this has no impact on the particle flux density or momentum transfer, while the energy transfer is reduced. Note, however, that all our conclusions are transferable to a quasi-2D situation with slightly reduced thermodynamic efficiencies due to the added energy transfer.

{\it Outward momentum and energy flux:}
On all of the walls shown in figure \ref{fig:schematic} the molecules are reflected diffusely. Correspondingly, we have
\begin{align}
\begin{Bmatrix}
\mathbf{F}_r(\mathbf{r}_{s}) \\ 
\varepsilon_r(\mathbf{r}_{s})
\end{Bmatrix} 
&=\int_{\mathbf{c}\cdot\mathbf{n}>0} (\mathbf{c}\cdot\mathbf{n})f_r(\mathbf{r}_s,\mathbf{c})
\begin{Bmatrix}
(m\mathbf{c}) \\ 
(\tfrac{1}{2} m\mathbf{c}^2)
\end{Bmatrix}
d^2\mathbf{c}
\\ \label{eq:outflux}
&=\left\{
\begin{array}{l}
m \nu(\mathbf{r}_{s})\left[ \frac{\sqrt{\pi}}{2} \bar{c}(\mathbf{r}_s) \,\mathbf{n} + u(\mathbf{r}_{s})\,\mathbf{t} \right] \\ 
\tfrac{1}{2} m \nu(\mathbf{r}_{s}) \left[ \frac{3}{2}\bar{c}^2(\mathbf{r}_s) + u^2(\mathbf{r}_s) \right].
\end{array}\right.
\end{align}
Here we have introduced the notation $\bar{c}=1/\sqrt{\beta}=\sqrt{2T/m}$ as a measure for the molecular velocity, corresponding to the most probable velocity of a molecule in the (three dimensional) Maxwell-Boltzmann distribution at temperature $T$ (not to be confused with the average or root-mean-square velocities).

Since on any specific wall segment the temperature and tangential velocity are constant, we use the notation $\bar{c}_i$ and $u_i$ for the molecular and wall velocities on segment $i$. Further we set $\hat{U}_i=u_i/\bar{c}_i$, i.e. we measure the wall tangential velocity in terms of the thermal velocity at the wall.

The first line of equation (\ref{eq:influx}) constitutes a Fredholm integral equation of the first kind for the particle flux density $\nu(\mathbf{r}_s)$ at all surfaces. Once the particle fluxes are known, heat and momentum fluxes on the wall can be calculated directly from equations (\ref{eq:influx}) and (\ref{eq:outflux}).

In this paper we solve the collisionless Boltzmann equation both by a discretisation of the Fredholm integral equation as well as with a Monte Carlo method, as described in the following two sections.

\section{Fredholm integral approach}
We split each wall into $N$ equally large segments, where segment $n$ is $\Omega_n^{(i)} = \left\{ \mathbf{r}_s^{(i)} \Big| (n-1)/N \le s \le n/N \right\}$. The integrals appearing in the first line of equation (\ref{eq:influx}) are approximated by \footnote{Due to the integral kernel of the form $\sim 1/r$, a direct approximation by a Riemann sum leads to increasing errors at corners. Since the particle flux $\nu$ itself is expected to remain finite and sufficiently smooth at these positions, we instead use the approximation detailed here.}
\begin{align}
I^{(ji)} &= \int_{\Omega^{(i)}} \frac{\cos \vartheta \cos \vartheta'}{r_{ss'}}\nu(\mathbf{r}_{s'}) G_{2}(\mathbf{r}_{s'},\vartheta') dl_{s'} \\
&\approx \sum_{n=1}^N \nu^{(i)}_n \int_{\Omega_n^{(i)}} \frac{\cos \vartheta \cos \vartheta'}{r_{ss'}} G_{2}(\mathbf{r}_{s'},\vartheta') dl_{s'}, \label{eq:intApprox}
\end{align}
where $\nu^{(i)}_n$ is a representative value for the particle flux emerging from segment $\Omega_n^{(i)}$ of line $i$. Due to the reflection symmetry of the problem, we have within our choice of parametrisation $\nu^{(\bar{1})}_n=\nu^{(1)}_n$ and $\nu^{(\bar{2})}_n=\nu^{(2)}_{N-n+1}$. In matrix notation, the Fredholm integral equation for the particle flux can be approximated as
\begin{align}\label{eq:FredholmDisc}
\begin{pmatrix}
\boldsymbol{\nu}^{(1)} \\ 
\boldsymbol{\nu}^{(2)}
\end{pmatrix} 
=
\begin{pmatrix}
\mathbf{K}^{(1\bar{1})} & \mathbf{K}^{(12)}+\mathbf{K}^{(1\bar{2})}\mathbf{T} \\ 
\mathbf{K}^{(21)}+\mathbf{K}^{(2\bar{1})} & \mathbf{K}^{(2\bar{2})}\mathbf{T}
\end{pmatrix} 
\begin{pmatrix}
\boldsymbol{\nu}^{(1)} \\ 
\boldsymbol{\nu}^{(2)}
\end{pmatrix},
\end{align}
where $T_{nm}=\delta_{n,N-m+1}$ and $\mathbf{K}^{(ij)}$ is the appropriate transfer matrix from wall $j$ to wall $i$ given in equation (\ref{eq:intApprox}). The Fredholm equation for the particle flux density is thus discretised to give $\boldsymbol{\nu}=\lambda \mathbf{K}\boldsymbol{\nu}$ with eigenvalue $\lambda=1$. Due to the approximation (\ref{eq:intApprox}) the spectrum of the matrix $\mathbf{K}$ will not exactly include the eigenvalue $\lambda=1$. However, it will contain a value very close to 1, clearly separated from the other eigenvalues with $|\lambda|<1$. The corresponding eigenvector $\boldsymbol{\nu}$ is the discretised particle flux density, from which momentum and energy transfer can be calculated.

\section{Test Particle Monte Carlo method}

In the collisionless regime the particle flux density at the boundary, $\nu$, and hence the full characterization of the system, can also be obtained by what is usually referred to as the {\em Test Particle Monte Carlo} (TPMC) method \citep{Duderstadt_1979, Bird_1994, Kersevan_2007}. Here a single particle's path is traced within the geometry, obeying the appropriate conditions at the boundaries, i.e. specular and diffuse reflection as well as periodic conditions. In the ergodic case (for example when sufficiently many diffuse walls are present \citep{Korsch_2002, Chumley_2013}), the distribution of reflection positions of the test particle gives the particle flux density $\nu$ in the limit of $N\to\infty$ reflections. Contrary to the {\em Direct Simulation Monte Carlo} method (DSMC, \cite{Bird_1994}) the velocity magnitude along each trajectory is unimportant for obtaining $\nu$. Moreover, considering the test particle as an ensemble of molecules encompassing the whole velocity spectrum, the same particle trajectory can be used to calculate momentum and energy transfer by weighing each collision with the appropriate moments of the velocity spectrum. Compared to DSMC, this results in faster convergence for momentum and energy transfer (although not for the particle flux density as mentioned above). Compared to the Fredholm integral method, needing convex domains with diffuse walls within our approach, this method is more versatile and very simple to implement, since any shadowing by walls is automatically taken care of by the routine identifying wall collisions. On the downside, the computational effort is much higher compared to the Fredholm integral approach. Details on the implementation of the TPMC can be found in appendix \ref{seq:TPMC_details}.

\section{Walls at rest} \label{sec:WallAtRest}
Before turning to the numerical evaluation, we would like to review and expand on some of the results obtained when assuming none of the walls are moving \citep{Donkov_2011}. In this case, the particle flux density $\nu(\mathbf{r})$ is constant \citep{Hardt_2009, Sone_2007} (see Sec.~2.5 of \cite{Sone_2007} for an elegant proof of this statement) and the integrals in equations (\ref{eq:influx}) and (\ref{eq:outflux}) can be evaluated analytically. Note in particular that they do not depend on the angle $\gamma$. For the flat wall 1 this is readily seen by noting that the angles $\vartheta$ under which walls at temperature $T_1$ or $T_2$ are seen (wall $\bar{1}$ and walls 2, $\bar{2}$, respectively) are independent of $\gamma$, as long as $\alpha>\gamma>\alpha-\pi/2$ \footnote{This even remains true when we replace wall 2 with an arbitrarily outward curved diffuse wall at constant temperature, i.e. a curve that has the same endpoints as the original one but never crosses the straight link between them into the domain.}. Due to momentum and energy conservation, this must also hold for the structured wall consisting of sections 2 and 3. By directly evaluating (\ref{eq:influxXi}) we obtain for the tangential and normal forces on the flat wall 1 \citep{Donkov_2011},
\begin{align}\label{eq:FtransStat}
F_{t}^{(1)}(u_1{=}0) &=  \frac{Lm\nu}{\sqrt{\pi}} \left(\bar{c}_2 - \bar{c}_1\right) \left(\frac{\pi}{2} - \alpha\right) \frac{\sin(2\alpha)}{2},\\
F_{n}^{(1)}(u_1{=}0) &= -\frac{\sqrt{\pi} Lm\nu}{2}\left(\bar{c}_2 + \bar{c}_1\right) + \frac{F_{t}^{(1)}(u_1{=}0)}{\tan\alpha},
\end{align}
and the transferred energy is
\begin{equation}\label{eq:Etrans}
\Delta\varepsilon_{12} (u_1{=}0) = \frac{3}{4}Lm\nu\left(\bar{c}_2^2 - \bar{c}_1^2\right)\sin\alpha.
\end{equation}
Note that as in \citep{Donkov_2011} these equations remain valid when the flat wall (1) does not directly rest on the structured one (2, 3), since for the calculation only the momentum and energy flux trough any parallel surface lying somewhere between the two is relevant. It is also easy to see that no tangential force is exerted on the top wall when wall 3 instead of being specular is diffuse and has the same temperature as wall 2. This is a simple consequence of the fact that particles arriving from direction $\vartheta$ at wall 1 have the same properties as particles seen under an angle $-\vartheta$, except that their tangential momentum is reversed since $f_i$ is the same in both cases, c.f. equation (\ref{eq:influxXi}). Moreover, we can generalize to a situation where wall 3 is partially diffuse (with probability $\tilde{\alpha}$) and partially specular (with probability $(1-\tilde{\alpha})$, where $\tilde{\alpha}$ is the accommodation coefficient), in which case the full solution is obtained by simple superposition.

For a diffusely reflecting wall at rest the tangential force is solely due to the impinging molecules, since the outgoing molecules are reflected symmetrically. On a moving wall the reflected particles will contribute with $-m\nu L u_1$ to the tangential force. For an order of magnitude assessment of the expected efficiency of our proposed device, let us assume that this is the only relevant effect due to the moving unstructured wall 1, i.e. we assume that the particle distribution $\nu$ along the walls is not strongly affected and neglect the additional momentum flux from backscattering of particles from the specularly reflecting wall. The harvested power $P=(F_{t}^{(1)}(u_1{=}0)-Lm\nu u_1)u_1$ thus becomes maximal for $u_1^\mathrm{max}=F_{t}^{(1)}(u_1{=}0)/(2Lm\nu)$ and is $P_\mathrm{max}=F_{t}^{(1)}(u_1{=}0)^2/(4Lm\nu)$. Using the same line of reasoning we approximate the transferred energy by equation (\ref{eq:Etrans}), and obtain as an estimate for the maximum efficiency, $\eta=P/\Delta\epsilon$, as function of the geometry parameters and wall temperatures
\begin{align}\label{eq:etaEstimate}
\eta_\mathrm{max}&=\frac{(\pi - 2 \alpha)^2 \cos^2(\alpha) \sin\alpha}{12\pi} \frac{|(1 - \bar{c}_1/\bar{c}_2)|}{(1 + \bar{c}_1/\bar{c}_2)}.
\end{align}
Note that this expression is symmetric under the exchange $\bar{c}_1 \rightleftharpoons \bar{c}_2$. Also note that for $\bar{c}_2/\bar{c}_1>1$ the tangential force $F_t^{(1)}>0$, and correspondingly the wall moves in direction $\mathbf{t}_1$ for the extraction of mechanical energy, while for $\bar{c}_2/\bar{c}_1<1$ it has to move in the direction of $-\mathbf{t}_1$. In this expression the thermal-velocity independent prefactor has a maximum at $\alpha\approx 22^\circ$ with value of $4.8\%$.

For the above estimate we assumed that in total the particles leaving the unstructured wall 1 carry away a tangential momentum of $Lm\nu u_1$. However, as noted above, in particular for small angles $\alpha$, much of this momentum is reflected back to wall 1, with hardly any change in its tangential component. In fact, the particle flux from wall 2 to wall 1 (and vice versa) is just $\nu L\sin\alpha$, as is readily seen when observing that for $\gamma=\alpha$ all particles leaving wall 2 eventually arrive at wall 1 (note that the expression for the transferred energy, equation (\ref{eq:Etrans}), takes this correctly into account). An alternative estimate for the contribution of the reflected particles to the tangential force on wall 1 is therefore $-m\nu (L\sin\alpha) u_1$. Retracing the steps leading to the efficiency estimate above and using the alternative expression for the force leads to $\tilde{\eta}_\mathrm{max}=\eta_\mathrm{max}/\sin\alpha$, which has its maximum at $\alpha\to 0$, where the thermal-velocity independent prefactor becomes $\pi/12\approx 26\%$.

We cannot stress enough that these simple estimates heavily rely on the assumption of constant particle flux density at the wall and only approximately take into account the backscattering from the specular wall 2. As it will turn out, some aspects of both estimates are recovered in different regimes. However, generally both overpredict the obtainable efficiency.

\section{Numerical Results}

We now turn to the numerical results obtained in the case of a moving wall. We will present the results normalized such that they are independent of the geometric length scale $L$ and the average particle flux density at the walls, as well as only implicitly dependent on the molecular mass $m$. The relevant parameters are the two geometric angles $\alpha$ and $\gamma$, the ratio of wall temperatures and the tangential velocity of the unstructured wall 1. As before, the latter two will be given in terms of the ratios involving the velocity scale $\bar{c}_i$ of a particle reflected diffusely from a wall at temperature $T_i$. Additionally, the gap size $H/L$ is relevant for results obtained within the TPMC method, while it vanishes in our implementation of the Fredholm integral method.

Unless explicitly stated otherwise, the results were obtained with $N=30$ grid points on each wall in the case of the Fredholm integral approach, and with $N=10^7$ boundary collisions in the case of the TPMC method. Under these conditions the computation times are roughly 20 times shorter with the Fredholm integral approach compared to the TPMC method. Unless error bars are explicitly displayed, discretisation errors and data scatter are estimated to be of the order of or smaller than the size of the data symbols used.

\begin{figure}
\includegraphics[width=8cm]{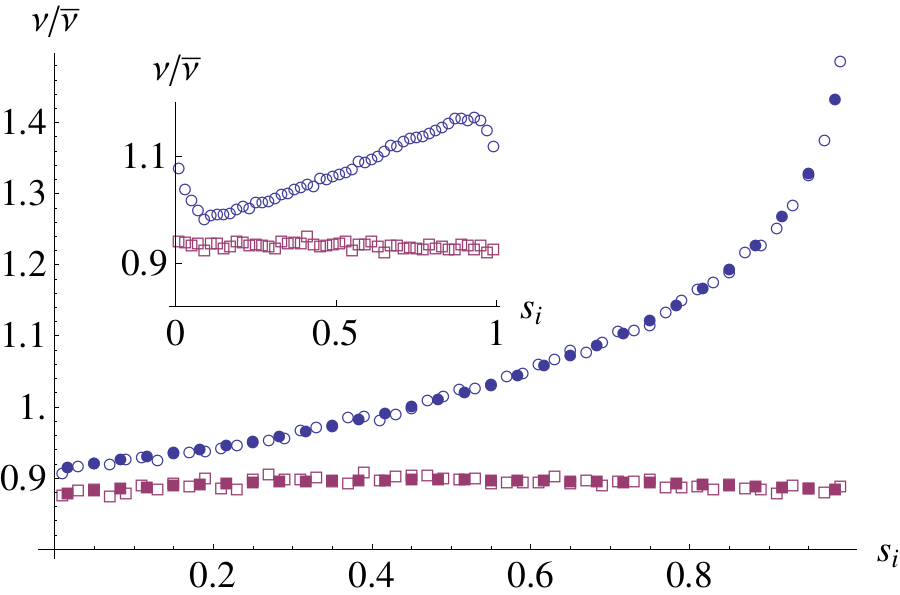}
\caption{\label{fig:nuUrel} (Color online) Particle flux density at wall 1 (upper curve, blue circles) and 2 (lower curve, red squares) for $\hat{U}_1=u_1/\bar{c}_1=0.1$, $\alpha=25^\circ$, $\gamma=0^\circ$ and $H/L=0$. Filled symbols were obtained using the Fredholm integral approach with $N=30$ grid points on each wall. Open symbols were obtained with the TPMC method, using $N=10^7$ wall collisions and evaluated using 50 bins along each wall. The inset shows the same situation with $H/L=0.1$ calculated with the TPMC method.
}
\end{figure}

Figure \ref{fig:nuUrel} shows the particle flux density $\nu$ along the diffuse boundaries, wall 1 and 2, with the unstructured wall 1 moving in direction of the tangent vector $\mathbf{t}_1$ shown in figure \ref{fig:schematic}. The flux density is normalized such that the integral of $\nu$ along the diffuse walls is unity. As mentioned before, its magnitude, $\bar{\nu}=\int\!\nu(\mathbf{r_s})\, dl_s$, drops out when considering force and energy ratios. In case of a wall at rest, the flux distribution is uniform, as dictated by the analytical result (not shown). As soon as the wall starts to move, particles become concentrated in the wedge region between wall 1 and the specular wall 3, as one would expect, and diluted at the opposite end. At the same time the flux density at the 'leeward' wall 2 is decreased but remains relatively homogeneous. Note that the particle flux density is not continuous at the edge $s_1=s_2=0$. For a wall moving in the opposite direction the distribution on wall 1 is essentially reversed, while the 'windward' wall 2 sees an increased particle flux. Note also that the integration kernel $G_{2}(\mathbf{r'}, \vartheta')$ is independent of the ratio of thermal velocities $\bar{c}_2/\bar{c}_1$, which therefore is also true for the particle flux densities.

Our Fredholm integral approach forces us to consider the idealized situation of a vanishing gap between the structured surface and the moving wall, i.e. $H/L=0$ in figure \ref{fig:schematic0}. This requirement can easily be relaxed within the TPMC method at the cost of a larger computational effort. The inset of figure \ref{fig:nuUrel} shows the particle flux density for $H/L=0.1$. As one would expect, the distribution on the unstructured wall 1 remains much flatter in this case and in particular the pile-up of particles in front of the ridges of the structure is not nearly as strong as in the case of a closed domain.

\begin{figure}
\includegraphics[width=8cm]{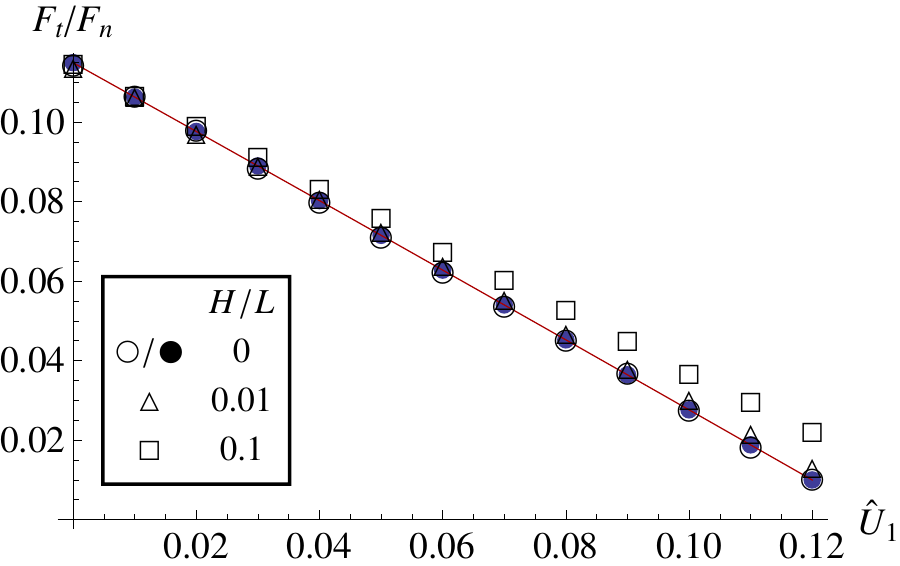}
\caption{\label{fig:ForceUrel} (Color online) Ratio between tangential and normal force on wall 1 as a function of wall velocity. $\bar{c}_2/\bar{c}_1=2$, $\alpha=25^\circ$, $\gamma=0^\circ$ and $H/L$ varying. The full circles correspond to values obtained with the Fredholm integral approach with $N=30$ grid points on each wall; the line is a linear fit to these data points. Open symbols are calculated within the TPMC method with $N=10^7$ boundary collisions for different separations $H/L$ of the moving wall from the structured wall.}
\end{figure}

\begin{figure}
\includegraphics[width=8cm]{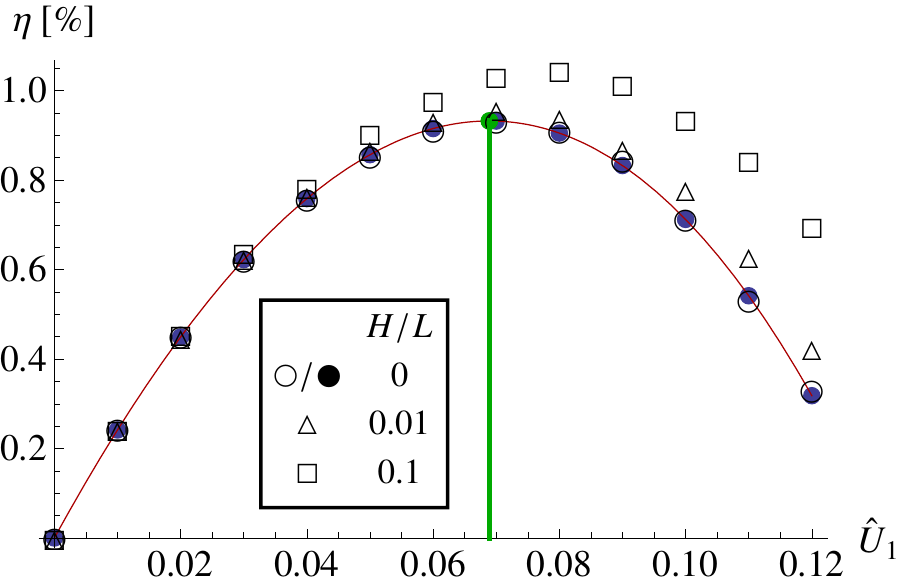}
\caption{\label{fig:EffUrel} (Color online) Efficiency of heat engine as a function of wall velocity. $\bar{c}_2/\bar{c}_1=2$, $\alpha=25^\circ$, $\gamma=0^\circ$ and $H/L$ varying. The full circles correspond to values obtained with the Fredholm integral approach with $N=30$ grid points on each wall. The (red) curve is a spline fit from which the maximum efficiency and corresponding wall velocity for the given geometry and wall temperatures is deduced (green vertical line). Open symbols are calculated within the TPMC method with $N=10^7$ boundary collisions for different separations $H/L$ of the moving wall from the structured wall.}
\end{figure}

Knowledge of the particle flux density allows calculating the forces on the moving wall 1, shown in figure \ref{fig:ForceUrel} for different relative wall velocities for a particular set of geometric parameters and wall temperatures. To a good approximation the tangential force decreases linearly with the wall velocity \footnote{We remark, that the deviation $(\nu/\bar{\nu}-1)$ of the particle flux density from the value at $\hat{U}_1=0$ is not linear in $\hat{U}_1$. Nevertheless, for small $\hat{U}_1$ this is approximately true, which together with the linearity of the outward momentum flux is the reason for the nearly linear decrease of the force in figure \ref{fig:ForceUrel}.}, which in turn means that the work extracted from the system initially increases but then goes through a maximum as the wall velocity is increased. This is reflected in figure \ref{fig:EffUrel} in terms of the thermodynamic efficiency, $\eta=F_t u_w/\Delta\varepsilon$. Note that the maximum efficiency occurs at relative velocities $\hat{U}_1{=}u_1/\bar{c}_1$ of the order of 0.1. Since at ambient temperatures the thermal velocities are several 100 m/s, this means that the wall has to move at a substantial speed. Both figures also show results obtained within the TPMC method (open symbols) in cases where $H/L\geq 0$. As expected, the reduced pile-up of particles in front of the ridges results in larger tangential forces in the case of a moving wall and hence larger efficiencies.

\begin{figure}
\includegraphics[width=8cm]{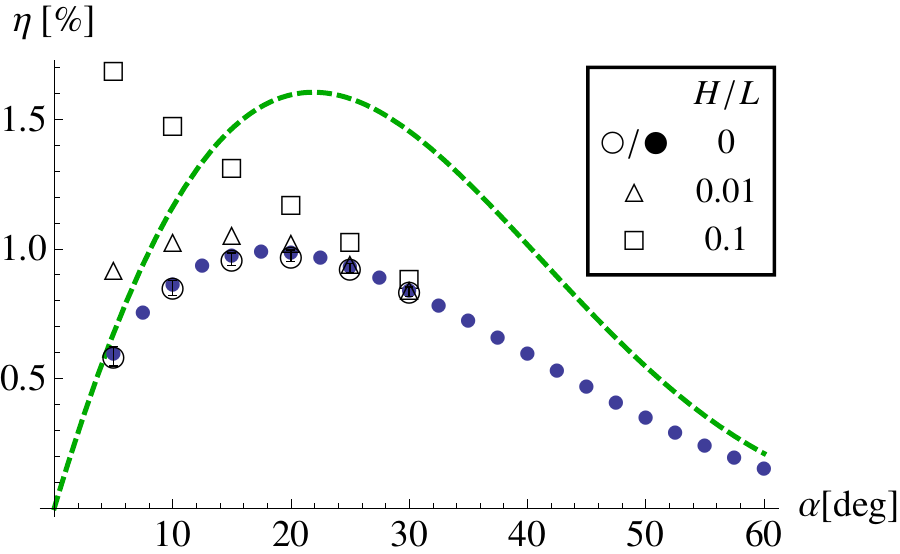}
\caption{\label{fig:EffAB2D} (Color online) Maximum efficiency of the heat engine as a function of $\alpha$. $\bar{c}_2/\bar{c}_1=2$, $\gamma=0^\circ$ and $H/L$ varying. The full circles correspond to values obtained with the Fredholm integral approach with $N=30$ grid points on each wall. Open symbols are calculated within the TPMC method with $N=10^7$ boundary collisions for different separations $H/L$ of the moving wall from the structured wall. The dashed green line corresponds to the efficiency estimate of equation (\ref{eq:etaEstimate}).}
\end{figure}

Repeating the procedure just outlined, we calculate the maximum efficiency as a function of the angle $\alpha$ as shown in figure \ref{fig:EffAB2D} for a thermal velocity ratio of $\bar{c}_2/\bar{c}_1=2$. Similar to our estimate (equation (\ref{eq:etaEstimate})) we find a strong dependence on $\alpha$. Within the Fredholm integral approach we have verified that the dependence on the angle $\gamma$ is very weak (with variations smaller than the symbol size if $-10^\circ \le \gamma \le 20^\circ$), in accordance with our estimate (equation (\ref{eq:etaEstimate})). The maximum efficiency is obtained at values of $\alpha$ roughly between 15$^\circ$ and 20$^\circ$ for $H/L=0$. In this case our estimate (equation (\ref{eq:etaEstimate})) is roughly a factor 1.5 higher than the calculated values. As expected, for $H/L>0$ the maximum efficiency  increases compared to the case $H/L=0$ due to the reduced accumulation of particles close to the ridges of the lower surface, c.f. figure \ref{fig:nuUrel}. It is interesting to note that for $H/L=0.1$ the maximum efficiency even increases above the estimate of equation (\ref{eq:etaEstimate}) for small angles. This can be attributed to direct backscattering for particles emitted from the flat wall 1 at the specular wall 2, as already noted in section \ref{sec:WallAtRest}. In that context we initially assumed that compared to the case of a moving wall all that happens is that each particle carries away an additional tangential momentum $m u_1$, reducing the tangential force. However, for small angles $\alpha$, this tangential momentum is only reduced a little upon specular reflection at wall 2, and much of it is returned to wall 1.

\begin{figure}
\includegraphics[width=8cm]{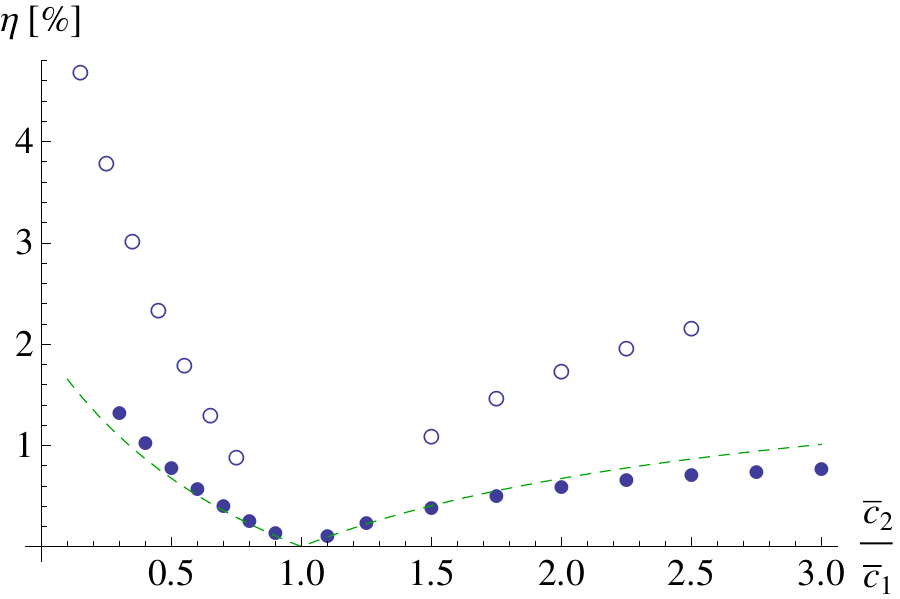}
\caption{\label{fig:EffUthermal} (Color online) Maximum efficiency of the heat engine as a function of the thermal velocity ratio, $\bar{c}_2/\bar{c}_1$, for $\alpha=5^\circ$, $\gamma=0^\circ$. The green line corresponds to the efficiency estimate, eq. (\ref{eq:etaEstimate}). Filled and open circles correspond to $H/L=0$ (evaluated with the Fredholm integral approach) and $H/L=0.1$ (using the TPMC method, $N=10^7$ collisions) respectively.}
\end{figure}

Finally, we investigate the dependence of the maximal efficiency on the ratio of thermal velocities $\bar{c}_2/\bar{c}_1$ in figure \ref{fig:EffUthermal} for shallow grooves ($\alpha=5^\circ$). We again compare the simple analytical estimate, eq. (\ref{eq:etaEstimate}), with numerical data obtained within the Fredholm integral approach ($H/L=0$) and the TPMC method ($H/L=0.1$). For $\bar{c}_2/\bar{c}_1>1$ wall 1 moves in direction $\mathbf{t}_1$ for the extraction of mechanical energy, and vice versa for $\bar{c}_2/\bar{c}_1<1$. As remarked previously, the analytical estimate is symmetric under the exchange $\bar{c}_1 \rightleftharpoons \bar{c}_2$. From the graph it can be seen that this is only approximately valid for the numerically obtained data. This is partially due to the different particle flux density distributions on the wall emerging in the two situations. Moreover, the symmetry between the two cases is broken by the fact that the wall velocity at maximum efficiency scales with the difference in thermal velocities, $\bar{c}_1-\bar{c}_2$, at least in our simple analytical estimate of section \ref{sec:WallAtRest}. Hence the relative wall velocity, $\hat{U}_1=u_1/\bar{c}_1$, which strongly influences the scattering behavior via the moments $G_n$ of the Maxwell-Boltzmann distribution, scales quite differently in the two cases. Nevertheless, the simple estimate is able to roughly reproduce the dependence of the efficiency on the thermal velocities at the walls.

Obviously, with increasing temperature ratio between the two walls the efficiency increases; according to the analytical estimate up to a maximum value dictated by the geometry.

\section{Alternative wall structures}\label{sec:AlternativeWallStructures}

\begin{figure}

\subfigure[]{
\begin{tikzpicture}[scale=2.75]
  \def\lenOne{1} 
  \def\lenH{0.3} 
  \def\lenW{0.125} 

  \coordinate (t1) at ($(0:1)$);   
  \coordinate (t2) at ($(90:1)$);

  \coordinate (a) at (0,0);
  \coordinate (b) at ($(a) - \lenH*(t2)$);
  \coordinate (c) at ($(b) + \lenOne*(t1)$);
  \coordinate (d) at ($(a) + \lenOne*(t1)$);
  
  \coordinate (e) at ($(a) + \lenW*(t2)$);
  \coordinate (f) at ($(e) + \lenOne*(t1)$);
  
  \draw[very thick,blue] (e) -- node[black,above] {$1$} (f);								
  \draw[very thick,color=BrickRed] (a) -- node[black,left] {$2$} (b);				
  \draw[very thick,gray] (b) -- node[black,below] {$3$} (c);								
  \draw[very thick,gray] (c) -- node[black,right] {$4$} (d);								
  
  \draw[very thick,gray,dotted] (a) -- node[black,left] {$P$} (e);
  \draw[very thick,gray,dotted] (d) -- node[black,right] {$P'$}  (f);  

\end{tikzpicture}
} 
\quad
%
\subfigure[]{
\begin{tikzpicture}[scale=1.4]
  \def\angBeta{30}
  \def\lenOne{1} 
  \def\lenR{1.5}
  \def\lenD{0.2}
  \def\lenW{0.3}

  \coordinate (t1) at ($(0:1)$);   
  \coordinate (t2) at ($(180+\angBeta:1)$); 
  \coordinate (t3) at ($(270+\angBeta:1)$); 
  \coordinate (t4) at ($(90:1)$);

  \coordinate (a) at (0,0);
  \coordinate (b) at ($(a) + 1/2*\lenOne*(t1) + \lenR*(t2) - 1/2*\lenD*(t3)$);
  \coordinate (c) at ($(a) + 1/2*\lenOne*(t1) + \lenR*(t2) + 1/2*\lenD*(t3)$);
  \coordinate (d) at ($(a) + \lenOne*(t1)$);
  
  \coordinate (e) at ($(a) + \lenW*(t4)$);
  \coordinate (f) at ($(e) + \lenOne*(t1)$);
  
  \coordinate (m1) at ($(a) + 1/2*\lenOne*(t1)$);                        
  \coordinate (m2) at ($(a) + 1/2*\lenOne*(t1) + \lenR*(t2)$);    
  \coordinate (mm1) at ($(a) + 1/4*\lenOne*(t1)$);                     
  \coordinate (mm2) at ($(a) + 1/2*\lenOne*(t1) + 1/2*\lenR*(t2)$);    

  \draw[very thick,blue] (e) -- node[black,above] {$1$} (f);								
  \draw[very thick,gray] (a) -- node[black,above left] {$3$} (b);						
  \draw[very thick,color=BrickRed] (b) -- node[black,below left] {$2$} (c);		
  \draw[very thick,gray] (c) -- node[black,below right] {$4$} (d);						
  
  \draw[very thick,gray,dotted] (a) -- node[black,left] {$P$} (e);
  \draw[very thick,gray,dotted] (d) -- node[black,right] {$P'$}  (f);
  
  
  \draw[gray,dashed] (m1) -- +(-180:0.45);
  \pgfmathparse{-180+\angBeta)}  																		
  \draw[gray,dashed] (m1) -- +(\pgfmathresult:0.45);
  \draw[] ($(m1)+(-180:0.3)$) node[below left] {$\alpha$} arc (-180:\pgfmathresult:0.3); 		
  
\end{tikzpicture}
} 

\caption{\label{fig:schematicGeneric} (Color online) Other generic geometries. Square grooves (a) and `ray guide' (b). Walls 1 and 2 are diffusely reflecting walls held at temperatures $T_1$ and $T_2$, respectively; walls 3 and 4 are specularly reflecting; dashed lines $P$ and $P'$ designate periodic pairs. In (b) $\alpha$ denotes the `inclination angle' of the guiding structure.}
\end{figure}
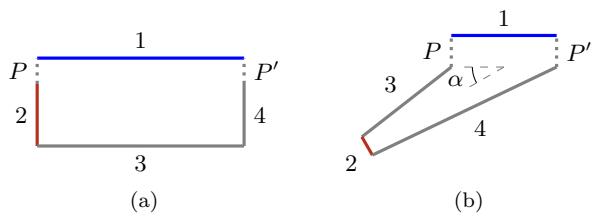

As we have seen, for the considered thermal velocity ratios the efficiency obtainable for the triangular configuration, figure \ref{fig:schematic0}, remains below roughly five percent, even under optimistic assumptions. The question arises to what extent this result is generic for a geometry of a structured and an unstructured surface at different temperature and whether we can do better. Besides the triangle another generic configuration of similar complexity is the square groove with two specular walls, c.f. figure \ref{fig:schematicGeneric} (a). From our intuition gained with the triangular geometry, we  expect that the highest efficiencies are obtained for shallow grooves. For both the triangular and square groove cases the specular wall essentially serves as a `guide' for particles leaving wall 2 towards wall 1 as well as a reflector for particles from wall 1. As such the efficiency obtained for the two cases is expected to be very similar.

Let us briefly elaborate on the idea of a guiding structure mentioned in the previous paragraph. Obviously, each particle moving at a velocity $c$ inevitably carries both momentum $mc$ and a kinetic energy of $mc^2/2$, and the velocity spectrum is dictated by the thermal Maxwell-Boltzmann statistics at the wall. Since energy and momentum transfer are inherently linked in this way, the best we can do for maximizing the force on the unstructured wall 1 due to particles leaving wall 2, the diffuse part of the structured wall, is to make these particles hit wall 1 at large angles with respect to its normal. Such a rectification of momentum can be achieved by a tapered trough with specular walls shown in figure \ref{fig:schematicGeneric} (b), which essentially serve as a `ray guide', directing particles from wall 2 and reflecting particles from wall 1. Qualitatively, this structure can be analyzed by multiple reflection of the geometry at the specular walls, similarly as we have done for the triangular geometry. Unfortunately, this procedure reveals that for small inclination angles $\alpha$ of the guiding structure the effect this geometry produces is qualitatively not much different from the guiding that the triangular structure already provides at small angles. Essentially, the forces due to incoming particles on a specific surface are determined by the angle under which other surfaces of given temperature are seen from that wall, which includes mirror images due to specular reflection. Therefore the forces become largest when a hotter (or colder) surface is seen under a small angle only, a situation already achieved with the shallow triangle. A more detailed discussion, together with some numerical calculations for this structure, can be found in appendix \ref{sec:rayGuide}. As an afterthought, we mention that with our current methods we are unable to analyze a situation with two structured walls moving relative to each other. It is possible that in such a situation a higher efficiency can be reached.

\section{Conclusion and Outlook}\label{sec:ConclusionOutlook}

To conclude, we have analyzed a new mechanism for conversion of thermal into mechanical energy relying on momentum transfer occurring in the free-molecular flow regime. It was found that with the considered device significant thermodynamic efficiencies should be achievable. Our analysis reveals that the geometry and temperature dependence of the obtainable efficiency can be estimated reasonably well by a simple analytical expression. The litmus test for the efficiency of a heat engine is of course a comparison with the Carnot efficiency of an ideal heat engine. For a Carnot cycle run between hot and cold reservoirs at $T_h$ and $T_c$, respectively, the efficiency is $\eta_C = 1-T_c/T_h = 1-\bar{c}_{T_c}^2/\bar{c}_{T_h}^2$ in our notation. Using our efficiency estimate (\ref{eq:etaEstimate}), the ratio of efficiencies roughly scales as $\eta_\mathrm{max}/\eta_C \sim M(\alpha)/(1+\bar{c}_{T_c}/\bar{c}_{T_h})^2$, where $M(\alpha)$ is the velocity independent prefactor in eq. (\ref{eq:etaEstimate}). For large temperature differences this ratio is dictated by $M(\alpha)$, which stays below 0.05 for all angles. It is thus mainly this geometrical factor that limits the possible energy extraction efficiency of the present system. Along with the paramount importance of the geometrical structure of the device comes the expectation that with more complex geometries, higher efficiencies will be achievable. Especially setups with two structured walls could be promising candidates. The analysis and optimization of such devices, however, requires considering changes of the domain boundaries over time, a task that is beyond the scope of the numerical methods employed here.

The analysis presented in this article can be applied to a gas at rarefied conditions. However, when considering a gas at standard pressure and temperature, the free-molecular flow regime we have focused on corresponds to a very small device dimension. At standard conditions, and taking into account the state-of-the-art of nanostructuring techniques, a Knudsen number of the order of one gives a more realistic scenario than free-molecular flow. This raises the question on how our results would be modified when considering the transition flow regime. To answer this question, one would have to solve the Boltzmann equation using an appropriate method such as DSMC. Since the velocity of the corresponding thermally-induced flow is very small compared to the molecular velocity, such simulations are computationally very expensive \cite{Donkov_2011}. The computational challenges become even more severe when parameter or optimization studies have to be conducted, as in the present article. For this reason we had decided to limit our studies to the free-molecular flow regime. 

To get a rough idea how the thermodynamic efficiency changes when going from the free-molecular flow to the transition flow regime, the following line of arguments can be employed. According to equation (\ref{eq:etaEstimate}), the maximum mechanical power scales approximately like $ P_\mathrm{max} \propto (F_{t}^{(1)})^2 $. From the DSMC simulations of ref.\ \cite{Donkov_2011} it is known that the tangential force reduces to about $2/3$ of the free-molecular flow value ($ \textrm{Kn} \rightarrow \infty $) when considering a Knudsen number of one. On the other hand, from Monte-Carlo simulations of heat transport in a thin nitrogen layer between two surfaces at fixed temperatures it can be deduced that the heat flux decreases to about $68 \%$ when reducing the Knudsen number from $10$ to $1$ \cite{Shan_2013}. In that case the Knudsen number was varied by increasing the distance between the parallel plates. When studying the transition between $ \textrm{Kn} \rightarrow \infty $ and $ \textrm{Kn} = 1 $, an even larger reduction factor is expected. Therefore, we find that upscaling the model domain to dimensions characteristic for a Knudsen number of one at standard conditions comes along with two different effects that roughly cancel each other when computing the thermodynamic efficiency: A decrease of the tangential force and a decrease of the heat flux. From these very simplistic arguments we would expect that the thermodynamic efficiency in the transition-flow regime is not much different from that in the free-molecular flow regime; however, in the continuum limit the efficiency must certainly vanish. Clearly, more quantitative studies based on a numerical solution of the Boltzmann equation are needed to answer these questions conclusively.

\begin{acknowledgments}
We thank S. Tiwari and A. Klar for many fruitful discussions. This work was supported by the German Research Foundation (DFG) through the Cluster of Excellence 259.
\end{acknowledgments}

\bibliography{litRareRatchet}



\balancecolsandclearpage

\appendix
\section*{Appendix}
\section{Moments of velocity, $G_{n}(\mathbf{r}_{s'}, \vartheta')$}\label{seq:G_n}

In this section we evaluate the function $G_{n}(\mathbf{r}_{s'}, \vartheta')$ defined in equation (\ref{eq:gn}). As specified in section \ref{sec:transfer}, we use $\mathbf{n}_{ss'}=(\mathbf{n}_{s'}\cos\vartheta'+\mathbf{t}_{s'}\sin\vartheta'$), where $\mathbf{n}_{s'}$ and $\mathbf{t}_{s'}$ are the normal and tangential unit vectors at position $\mathbf{r}_{s'}$ on the boundary. Implicitly, $G_{n}(\mathbf{r}_{s'}, \vartheta')$ depends on both $u(\mathbf{r}_{s'})$, the tangential wall velocity at $\mathbf{r}_{s'}$, and $\bar{c}(\mathbf{r}_{s'})$, the thermal velocity scale of a molecule reflected at $\mathbf{r}_{s'}$. For convenience, we introduce their ratio, $\hat{U}(\mathbf{r}_{s'})=u(\mathbf{r}_{s'})/\bar{c}(\mathbf{r}_{s'})$. Then
\begin{widetext}
\begin{align}
G_{n}(\mathbf{r}_{s'}, \vartheta')  &= \int_0^\infty c^n\, \mathsf{F}^{2D}(\mathbf{r}_{s'}, c\,\mathbf{n}_{ss'})\,dc \nonumber\\
&=\frac{e^{-\hat{U}(\mathbf{r}_{s'})^2} \bar{c}(\mathbf{r}_{s'})^{n-2}}{\sqrt{\pi }}
	\Bigg[ \Gamma\! \left(\frac{n+1}{2}\right) \, _1F_1\left(\frac{n+1}{2};\frac{1}{2};\left(\hat{U}(\mathbf{r}_{s'}) \sin\vartheta'\right)^2\right)   \label{eq:gn_analytic}\\
&\qquad\qquad\qquad\qquad+\left(\hat{U}(\mathbf{r}_{s'}) \sin\vartheta'\right) n\Gamma \left(\frac{n}{2}\right) \, _1F_1\left(\frac{n}{2}+1;\frac{3}{2};\left(\hat{U}(\mathbf{r}_{s'}) \sin\vartheta'\right)^2 \right)\Bigg]. \nonumber
\end{align}
\end{widetext}
Here $_1F_1(a; b; z)$ is the Kummer confluent hypergeometric function \cite{Abramowitz_1970}, which has the series representation $_1F_1(a; b; z)=\sum_{n=0}^\infty\frac{(a)_n z^n}{(b)_n n!}$, with $(c)_n=\Gamma(c+n)/\Gamma(c)$ being the rising factorial (or Pochhammer function), $(c)_0=1$, $(c)_n=c(c+1)(c+2)\cdots(c+n-1)$.

More familiar forms can be obtained by expanding $G_{n}(\mathbf{r}_{s'}, \vartheta')$ in $\hat{U}(\mathbf{r}_{s'})$:
\begin{align}
G_{n}(\mathbf{r}_{s'}, \vartheta')  &\approx \nonumber\\
\label{eq:gnSeries}
&\hspace{-1.1cm} 
\left\{\begin{array}{ll}
\frac{1}{2} + \sqrt{\frac{4}{\pi}} \hat{U}(\mathbf{r}_{s'}) \sin\vartheta' & \text{for } n=2\\ 
\bar{c}(\mathbf{r}_{s'}) \left( \frac{1}{\sqrt{\pi}}+\frac{3}{2} \hat{U}(\mathbf{r}_{s'}) \sin\vartheta' \right) & \text{for } n=3\\ 
\bar{c}(\mathbf{r}_{s'})^2\Big( \frac{3}{4}+\frac{4}{\sqrt{\pi}} \hat{U}(\mathbf{r}_{s'}) \sin\vartheta'  &  \\
\quad\quad\; -\frac{3}{4}\left(\hat{U}(\mathbf{r}_{s'})\right)^2\left(1-5\sin^2\vartheta'\right) \Big) & \text{for } n=4. \\
\end{array} \right.
\end{align}
Note that since the lowest order correction to the energy transfer scales as ${\sim}u_i^2$, the transfer function $G_4$ needs to be expanded up to second order in $\hat{U}(\mathbf{r}_{s'})$. For our numerical calculations within the Fredholm integral approach, we use the analytical form (\ref{eq:gn_analytic}), which can be simplified for the specific values of $n$. However, for the TPMC method we adopt the series expansion (\ref{eq:gnSeries}) for calculating the momentum and energy transfer in order to reduce the computational effort. Therefore we limit the relative wall velocity to $\hat{U}\lesssim 0.1$ for all calculations using the TPMC method (which, as it turns out, is not a strong restriction).

\section{Details on the test particle Monte Carlo method}\label{seq:TPMC_details}

As mentioned in the main text, the test particle Monte Carlo method is easily implemented and consists of a straight-line ray-tracing routine for movement from one boundary to the next and an implementation of the boundary conditions, specifying the trajectory after collision with a boundary. 

As boundary conditions we implement diffuse and specular reflection as well as periodicity. Under specular reflection the normal component of a particle's momentum is reversed, while the component tangential to the boundary is conserved. Periodic boundaries result in a translation of the particle position while the momentum is conserved. For diffuse reflection, we note that the normalized particle flux density $\mathsf{P}_\mathbf{r}(\mathbf{c}) \equiv (\mathbf{c}\cdot\mathbf{n})f_r(\mathbf{r},\mathbf{c})/\nu(\mathbf{r})$ can be interpreted as the probability of a particle being emitted with velocity $\mathbf{c}$ from the wall. This can be written as a product of the probability densities for the normal and tangential velocity components (with $\mathbf{c}=c_n\mathbf{n}+c_t\mathbf{t}$; $c_n\in\mathbb{R}^+$, $c_t\in \mathbb{R}$)
\begin{align}
\mathsf{P}_\mathbf{r}(\mathbf{c}) &= c_n \mathsf{F}^{2D}(\mathbf{r}, \mathbf{c}) = \mathsf{P}_\mathbf{r}^{n}(c_n) \mathsf{P}^{t}_\mathbf{r}(c_t), \label{eq:velDistDiffuseProduct}\\
\mathsf{P}^{n}_\mathbf{r}(c_n) &= 2\beta c_n e^{-\beta c_n^2}, \label{eq:velDistNormal}\\
\mathsf{P}^{t}_\mathbf{r}(c_t) &= \sqrt{\beta/\pi}\, e^{-\beta (c_t-u)^2}, \label{eq:velDistTang}
\end{align}
where the wall moves at velocity $u$ in tangential direction, and $\mathbf{n}$, $\mathbf{t}$ are the unit wall normal and tangential vectors at position $\mathbf{r}$. Based on the inverse transformation sampling method (inversion of the cumulative distribution function, \cite{Press_1992}) and on the Box-Muller transform for the tangential component \cite{Press_1992}, efficient algorithms exist for generating velocities in accordance with these distributions (Weibull and normal). Following \cite{Shen_2006}, the random variables $C_n = \sqrt{-\ln X}/\sqrt{\beta}$ and $C_t = (\sqrt{-\ln Y}\cos(2\pi Z)/\sqrt{\beta}+u)$ are distributed according to equations (\ref{eq:velDistNormal}) and (\ref{eq:velDistTang}) when $X$, $Y$ and $Z$ are random variables uniformly distributed in the interval $[0,1]$.

For the TPMC method only the normalized velocity vector, $\hat{\mathbf{c}}=\mathbf{c}/c$, matters. In order to calculate the momentum and energy each ray carries away under the angle $\vartheta=\arcsin(c_t/c)$ from a diffuse wall, the moments $\langle c^n \rangle_{\mathbf{r}} (\vartheta)$ are needed. In terms of the angle and velocity magnitude, $\vartheta$ and $c$, the scattering kernel corresponding to equation (\ref{eq:velDistDiffuseProduct}) becomes (with $\mathbf{c}(\vartheta, c)=(\mathbf{n}\cos\vartheta+\mathbf{t}\sin\vartheta)c$; $\vartheta\in [-\pi,\pi]$, $c\in\mathbb{R}^+$)
\begin{equation}\label{eq:diffuseScatteringKernel}
p_\mathbf{r}(\vartheta,c)=\cos\vartheta\, c^2\, \mathsf{F}^{2D}(\mathbf{r}, \mathbf{c}(\vartheta, c)),
\end{equation}
which is the probability of a particle being emitted into angles between $\vartheta$ and $\vartheta+d\vartheta$ with respect to the wall normal and with velocity magnitudes between $c$ and $c+dc$. Thus
\begin{align}
\langle c^n \rangle_{\mathbf{r}} (\vartheta) &= \int_0^\infty \! c^n\, p_{\mathbf{r}}(c|\vartheta)\,dc = \int_0^\infty \! c^n\, \frac{p_{\mathbf{r}}(\vartheta,c) }{p_{\mathbf{r}}(\vartheta)}\,dc\\
&= \frac{\cos\vartheta}{p_{\mathbf{r}}(\vartheta)} \int_0^\infty \! c^{n+2}\,\mathsf{F}^{2D}(\mathbf{r}, \mathbf{c}(\vartheta, c))\,dc,
\end{align}
where we have used the definition of the conditional probability $p_{\mathbf{r}}(c|\vartheta)=p_{\mathbf{r}}(\vartheta,c) / p_{\mathbf{r}}(\vartheta)$ and marginal probability $p_\mathbf{r}(\vartheta)=\int_0^\infty\!\! p_\mathbf{r}(\vartheta, c)\,dc$. Since $\langle c^0 \rangle_{\mathbf{r}} (\vartheta) =1$ and with the definition (compare with equations (\ref{eq:gn}) and (\ref{eq:gn_analytic}))
\begin{equation}
G_{n}(\mathbf{r}, \vartheta) = \int_0^\infty\! c^{n}\,\mathsf{F}^{2D}(\mathbf{r}, \mathbf{c}(\vartheta, c))\,dc,
\end{equation}
we further get $p_{\mathbf{r}}(\vartheta) = \cos\vartheta\,G_{\mathbf{r}, 2}(\vartheta)$ and finally
\begin{equation}
\langle c^n \rangle_{\mathbf{r}} (\vartheta)  = \frac{G_{n+2}(\mathbf{r}, \vartheta)}{G_{2} (\mathbf{r}, \vartheta)}.
\end{equation}
Note that these moments are conserved on specular reflection and at periodic boundaries and thus will simply be carried further along to the next diffuse boundary. Also note that restricting the attention to rays, i.e. only velocity magnitudes play a role during tracing, only stationary states can be simulated. In particular no normal wall movement is allowed since that changes the geometry. 

Finally, we note that for a stationary wall the probability distribution of scattering angles is Lambert's law, $p_\mathbf{r}(\vartheta)=\cos(\vartheta)/2$. This can be effectively sampled using the inverse transformation sampling method, i.e. the random variable $\Theta = \arcsin X$ will be distributed according to Lambert's law when $X$ is a random variable uniformly distributed in the interval $[-1,1]$. Unfortunately, in case of a moving wall no analytic inverse of the cumulative distribution function of $p_\mathbf{r}(\vartheta)$ is known and the inverse transformation sampling method can only be used approximately, e.g. by a series expansion of $p_\mathbf{r}(\vartheta)$ in the small parameter $\hat{U}(\mathbf{r})$. However, the computational cost of this method turns out to be high (already Lambert's law requires the inverse of a trigonometric function) and the method of independently sampling a normal and tangential velocity component described above is faster and more exact.

To complete the discussion of the Monte Carlo approach we remark that in the limit of infinitely many traced rays, $N\to\infty$, each ray in the whole set of rays can be thought of as being distributed according to the probability density
\begin{equation}
p(\mathbf{r}, \vartheta, c) = p(\mathbf{r}) p(\vartheta, c | \mathbf{r}),
\end{equation}
for rays originating at some point $\mathbf{r}$ on the diffuse boundary. The conditional distribution $p(\vartheta, c|\mathbf{r})$ is given by equation (\ref{eq:diffuseScatteringKernel}) while the $\mathbf{r}$-distribution is generated by the ray tracing method in the limit of $N\to\infty$
\begin{equation}
p(\mathbf{r})\,dr \sim \nu(\mathbf{r})\,dr \sim N_{dr}(\mathbf{r})/N,
\end{equation}
where $N_{dr}(\mathbf{r})$ is the number of wall collisions within a region of width $dr$ around $\mathbf{r}$, and $\nu(\mathbf{r})$ is the particle flux density.

Due to the stochastic nature of the sampling method in TPMC, with $N$ wall collisions the convergence is only ${\sim 1/\sqrt{N}}$, potentially requiring a large number of collisions. In the Fredholm integral approach with $N$ bins on each wall the convergence is ${\sim 1/N}$ while the effort rises ${\sim N^2}$, scaling even worse than the Monte Carlo method. However, as seen on figure \ref{fig:nuUrel}, a smooth wall distribution is easily obtained for $N=30$ grid points on each wall within the integral approach, while a comparably smooth distribution takes $N\sim 10^7$ collisions for the TPMC.

\section{`Ray Guide' geometry} \label{sec:rayGuide}

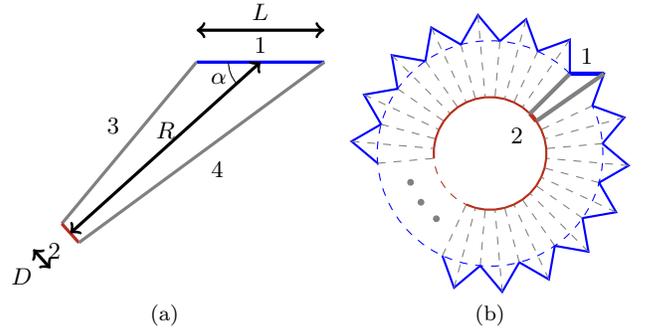
\begin{figure}

\subfigure[]{
\begin{tikzpicture}[scale=1.7]
  \def\angBeta{42}
  \def\lenOne{1} 
  \def\lenR{2.}
  \def\lenD{0.2}

  \coordinate (t1) at ($(0:1)$);   
  \coordinate (t2) at ($(180+\angBeta:1)$);
  \coordinate (t3) at ($(270+\angBeta:1)$);
  \coordinate (t4) at ($(90:1)$);

  \coordinate (a) at (0,0);
  \coordinate (b) at ($(a) + 1/2*\lenOne*(t1) + \lenR*(t2) - 1/2*\lenD*(t3)$);
  \coordinate (c) at ($(a) + 1/2*\lenOne*(t1) + \lenR*(t2) + 1/2*\lenD*(t3)$);
  \coordinate (d) at ($(a) + \lenOne*(t1)$);
  
  \coordinate (m1) at ($(a) + 1/2*\lenOne*(t1)$);   
  \coordinate (m2) at ($(a) + 1/2*\lenOne*(t1) + \lenR*(t2)$);

  \draw[very thick,blue] (a) -- node[black,above] {$1$} (d);								
  \draw[very thick,gray] (a) -- node[black,above left] {$3$} (b);						
  \draw[very thick,color=BrickRed] (b) -- node[black,below left] {$2$} (c);		
  \draw[very thick,gray] (c) -- node[black,below right] {$4$} (d);						
  

  \pgfmathparse{-180+\angBeta)}  																	
  \draw ($(m1)+(-180:0.25)$) arc (-180:\pgfmathresult:0.25);						
  \draw ($(m1)+(-180+\angBeta/2:0.35)$) node {$\alpha$};

  \draw[<->, very thick] ($(a)+0.25*(t4)$) -- node[black,above] {$L$} ($(d)+0.25*(t4)$);
  \draw[<->, very thick] ($(b)+0.3*(t2)$) -- node[black,below left] {$D$} ($(c)+0.3*(t2)$);
  \draw[<->, very thick] (m1) -- node[black,above] {$R$} (m2);

\end{tikzpicture}
} 
\subfigure[]{
\begin{tikzpicture}[scale=0.15]
  \def\angA{45} 
  \def\lenR{10} 
  \def\lenRi{5} 
  \def\angG{10} 

  \pgfmathparse{\lenR*sin(\angA)/sin(\angA-\angG)}  
  \def\lenRo{\pgfmathresult} 

  \coordinate (x0) at ($(\angA:\lenR)$);   
  \coordinate (y0) at ($(\angA-\angG:\lenRo)$);
  \coordinate (xi0) at ($(\angA:\lenRi)$);   
  \coordinate (yi0) at ($(\angA-\angG:\lenRi)$);

  \foreach \i in {1,2,3,4,5,6,7,8} {
      \coordinate (x\i) at ($(\angA+2*\angG*\i:\lenR)$);
      \coordinate (xm\i) at ($(\angA-2*\angG*\i:\lenR)$);
      \coordinate (y\i) at ($(\angA-\angG+2*\angG*\i:\lenRo)$);
      \coordinate (ym\i) at ($(\angA-\angG-2*\angG*\i:\lenRo)$);
      
      \coordinate (xi\i) at ($(\angA+2*\angG*\i:\lenRi)$);
      \coordinate (xim\i) at ($(\angA-2*\angG*\i:\lenRi)$);
      \coordinate (yi\i) at ($(\angA-\angG+2*\angG*\i:\lenRi)$);
      \coordinate (yim\i) at ($(\angA-\angG-2*\angG*\i:\lenRi)$);
  };

  \draw[thick, blue] (xm8) -- (ym7) -- (xm7) -- (ym6) -- (xm6) -- (ym5) -- (xm5) -- (ym4) -- (xm4) -- (ym3) -- (xm3) -- (ym2) -- (xm2) -- (ym1) -- (xm1) -- (y0) -- (x0) -- (y1) -- (x1) -- (y2) -- (x2) -- (y3) -- (x3) -- (y4) -- (x4) -- (y5) -- (x5) -- (y6) -- (x6) -- (y7) -- (x7);
  \draw[thick, color=BrickRed] (xim8) -- (yim7) -- (xim7) -- (yim6) -- (xim6) -- (yim5) -- (xim5) -- (yim4) -- (xim4) -- (yim3) -- (xim3) -- (yim2) -- (xim2) -- (yim1) -- (xim1) -- (yi0) -- (xi0) -- (yi1) -- (xi1) -- (yi2) -- (xi2) -- (yi3) -- (xi3) -- (yi4) -- (xi4) -- (yi5) -- (xi5) -- (yi6) -- (xi6) -- (yi7) -- (xi7);
  
  \foreach \i in {1,2,3,4,5,6,7} {
      \draw[gray,dashed] (x\i) -- (xi\i);
      \draw[gray,dashed] (y\i) -- (yi\i);
      \draw[gray,dashed] (xm\i) -- (xim\i);
      \draw[gray,dashed] (ym\i) -- (yim\i);
  };
  \draw[gray,dashed] (xm8) -- (xim8);

  \pgfmathparse{(\lenR+\lenRi)/2}
  \def\lenRmid{\pgfmathresult} 
  \coordinate (xmid) at ($(200:\lenRmid)$);
  \draw[decorate,decoration={shape backgrounds,shape=circle,shape size=0.75mm,shape sep=3mm},fill,gray] (xmid) arc (200:235:\lenRmid); 
  
  \draw[ultra thick,color=BrickRed] (xi0) -- node[black,below left]{$2$} (yi0);
  \draw[ultra thick,blue] (x0) -- node[black,above]{$1$} (y0);
  \draw[ultra thick,gray] (xi0) -- (x0);
  \draw[ultra thick,gray] (yi0) -- (y0);

  \pgfmathparse{360+\angA-2*\angG)}  
  \draw[blue,dashed] (x0) arc (\angA:\pgfmathresult:\lenR);
  \draw[color=BrickRed,dashed] (\lenRi,0) arc (0:360:\lenRi);
  
\end{tikzpicture}
} 
\caption{\label{fig:schematicGunCombined} (Color online) (a) The `ray guide' geometry is parametrised by the angle $\alpha$ and the three lengths $L$, $R$ and $D$. Walls 1 and 2 are diffusely reflecting, being held at temperatures $T_1$ and $T_2$, respectively. Wall 3 and 4 are specularly reflecting walls. (b) Rosette obtained by repeatedly mirroring the `ray guide' geometry at the specular walls 3 and 4.}
\end{figure}

In section \ref{sec:AlternativeWallStructures} we surmised that an optimally efficient heat engine should eject molecules from a hot towards a cold surface in such a way that the particle's momentum is absorbed mainly tangentially to the receiving surface. A promising geometry to accomplish this is the wedge shaped `ray guide' shown in figure \ref{fig:schematicGunCombined}(a). Here any molecule leaving the diffuse wall 2 is guided between the two specular walls 3 and 4 towards the second diffuse wall 1. Due to the tapering of the wedge the momentum of the molecules will be aligned with the wedge (along the double-sided arrow in figure \ref{fig:schematicGunCombined}(a)), and so the momentum transfer from wall 2 to wall 1 can be made to occur almost tangentially to wall 1. Similarly, particles leaving wall 1 into the trough have a high probability of being reflected unless they are aimed almost directly at wall 2. If for the sake of the argument we assume for the moment that $T_1\ll T_2$, i.e. the energy and momentum leaving wall 1 can be neglected, we might hope to have a suitable `momentum rectifier'.

In order to qualitatively analyze this system we proceed similarly as for the wedge-shaped geometry. First we mirror the geometry repeatedly along the specular walls to obtain the `rosette' shown in figure \ref{fig:schematicGunCombined}(b). Since we are interested in net momentum and energy transfer at wall 1, which can be calculated via equation (\ref{eq:influxXi}), it is not necessary to complete the rosette, since it is enough knowing the temperature (and particle flux density) at the wall under a particular line of sight in order to calculate the transfer. When all walls are at rest the particle flux density on all walls is constant, and it is not too hard to calculate the force on wall 1. 

However, let us continue the qualitative analysis with $T_1\ll T_2$. From figure \ref{fig:schematicGunCombined}(b) we deduce that the force becomes large, when the `inner polygon' occupies much of the viewing angles to the left, while the viewing angles to the right are shielded by the `outer polygon' (blue curve). Such a situation is achieved for small angles $\alpha$ and a sufficiently large `inner polygon'. However, this situation is qualitatively not much different from the situation in our triangular geometry for small opening angles. It is thus expected that the forces and efficiencies will not greatly deviate from the ones we have found in the detailed analysis of the triangular geometry. A similar reasoning can be performed for the opposite case of $T_1\gg T_2$, where now we would want $2\alpha\approx\pi/2$, and again a large `inner polygon' such that the right field of view is largely occupied by the hot wall while the left field of view is `cold'.

\begin{figure}
\null\hfill\subfigure[]{\includegraphics[width=4cm]{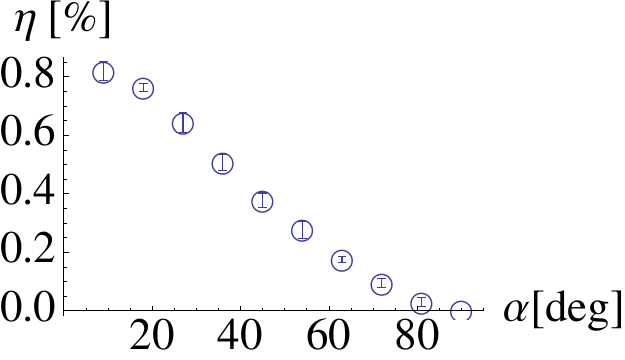}}\hfill
\subfigure[]{\includegraphics[width=4cm]{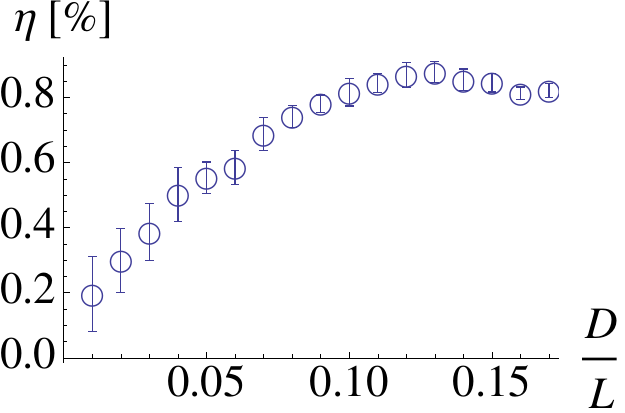}}\hfill\null\\
\subfigure[]{\includegraphics[width=4cm]{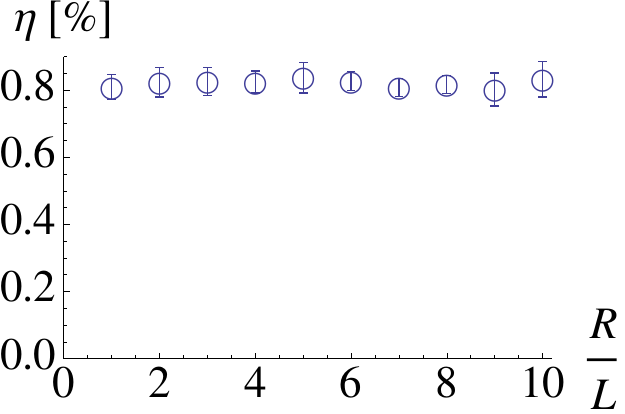}}
\caption{\label{fig:RayGunEff} (Color online) Maximum efficiency of the ``ray guide'' heat engine as a function of the geometry parameters for $\alpha=9^\circ$. $\bar{c}_2/\bar{c}_1=2$. The base parameter set is $R/L=4$, $D/L=0.1$. Results using the TPMC method with $N=10^7$ boundary collisions.}
\end{figure}

In order to underpin this simple qualitative argument, we have calculated the efficiencies that can be obtained using this `ray guide' geometry for different angles and lengths $D/L$ and $R/L$, c.f. figure \ref{fig:RayGunEff}. To be able to compare with the triangular geometry we again use the ratio $\bar{c}_2/\bar{c}_1 = 2$. As expected, varying $\alpha$ at fixed $D/L=0.1$ and $R/L=4$, we see that the efficiency decreases drastically with larger angles, illustrating the momentum-rectifying nature of the geometry (figure \ref{fig:RayGunEff}(a)). Next, using a small angle $\alpha=9^\circ$, we vary $D/L$, illustrating the importance of a large `inner polygon' in the rosette of figure \ref{fig:schematicGunCombined}(b). Note that the efficiency approximately saturates for $D/L\approx 0.1$, close to $D/L=\sin\alpha\approx 0.16$, where wall 2 becomes the projection of wall 1 under the angle $\alpha$ (figure \ref{fig:RayGunEff}(b)). Due to this fact the efficiency does not strongly depend on $R/L$ for this relatively large value of $D/L=0.1$ (figure \ref{fig:RayGunEff}(c)). This can be explained by noting that in figure \ref{fig:schematicGunCombined}(b) the `radius' of the `inner polygon' scales approximately as $R_D/R\approx(L\sin\alpha/D-1)^{-1}$ for not too large tapering angles, showing that while changing the distance of the inner polygon its size is scaled accordingly. Thus the angle under which the 'inner polygon' is seen from wall 1 remains the same.

As expected from our qualitative analysis, the efficiency remains at the same order of magnitude as for the triangular geometry, analyzed throughout the main part of the paper.

\end{document}